\documentclass{emulateapj}
\usepackage{epsfig}
\usepackage{graphicx}

\def\ltsima{$\; \buildrel < \over \sim \;$}
\def\lsim{\lower.5ex\hbox{\ltsima}}
\def\gtsima{$\; \buildrel > \over \sim \;$}
\def\gsim{\lower.5ex\hbox{\gtsima}}

\shorttitle{The Rate of CC Supernovae}
\shortauthors{Dahlen et al.}

\begin{document}


\title{The Extended {\it Hubble Space Telescope} Supernova Survey: The Rate of Core Collapse Supernovae to z$\sim$1}


\author{Tomas Dahlen\altaffilmark{1}, 
Louis-Gregory Strolger\altaffilmark{2},
Adam G. Riess\altaffilmark{1,3},
Seppo Mattila\altaffilmark{4},
Erkki Kankare\altaffilmark{4},
and Bahram Mobasher\altaffilmark{5}
}
\email{dahlen@stsci.edu}

\altaffiltext{1}{Space Telescope Science Institute, 3700 San Martin Drive, Baltimore, MD 21218, USA}
\altaffiltext{2}{Department of Physics and Astronomy, Western Kentucky University, Bowling Green, KY 42101, USA}
\altaffiltext{3}{Department of Physics and Astronomy, Johns Hopkins University, 3400 North Charles Street, Baltimore, MD 21218, USA}
\altaffiltext{4}{Tuorla Observatory, Department of Physics and Astronomy, University of Turku, V\"{a}is\"{a}l\"{a}ntie 20, FI-21500 Piikki\"{o}, Finland}
\altaffiltext{5}{Department of Physics and Astronomy, University of California, Riverside, CA 92521, USA}
\begin{abstract}
We use a sample of 45 core collapse supernovae detected with the Advanced Camera for Surveys on board the {\it Hubble Space Telescope} to derive the core collapse supernova rate in the redshift range $0.1<z<1.3$. In redshift bins centered on $\langle z\rangle=0.39$,  $\langle z\rangle=0.73$, and   $\langle z\rangle=1.11$, we find rates 3.00$^{+1.28}_{-0.94}$$^{+1.04}_{-0.57}$, 7.39$^{+1.86}_{-1.52}$$^{+3.20}_{-1.60}$, and 9.57$^{+3.76}_{-2.80}$$^{+4.96}_{-2.80}$, respectively, given in units of  yr$^{-1}$Mpc$^{-3}~10^{-4}~h_{70}^3$.  The rates have been corrected for host galaxy extinction, including supernovae missed in highly dust-enshrouded environments in infrared bright galaxies. The first errors are statistical while the second ones are the estimated systematic errors. We perform a detailed discussion of possible sources of systematic errors and note that these start to dominate over statistical errors at $z>0.5$, emphasizing the need to better control the systematic effects. For example, a better understanding of the amount of dust extinction in the host galaxies and knowledge of the supernova luminosity function, in particular the fraction of faint $M \gsim-15$~supernovae, is needed to better constrain the rates. When comparing our results with the core collapse supernova rate based on the star formation rate, we find a good agreement, consistent with the supernova rate following the star formation rate, as expected.

\end{abstract}


\section{Introduction}
Supernovae (SNe) mark the end of the life cycle of certain stars. Studies of SNe are important both for understanding the physics leading to and driving these explosions, as well as the impact that the SNe have on their environments. As the main producers of heavy elements, SNe are pivotal for understanding the chemical evolution in galaxies as well as the intergalactic medium via SN-driven outflows from galaxies. SNe are also thought to be one of the producers of dust in the universe (e.g., review by Kozasa et al. 2009). Furthermore, SNe tie into feedback processes regulating galaxy formation. To understand how these processes evolve with cosmic time, it is important to understand how the number of exploding SNe changes with redshift, i.e., the evolution of the cosmic SN rate (SNR). Since SNe, particularly the core collapse (CC) SNe, have massive progenitors with short main-sequence lifetimes, the rate should closely follow the star formation rate (SFR), offering a direct and independent way of measuring the SFR and at the same time the metal enrichment rate. While this method is straightforward in principle, it is hampered particularly by the need for corrections due to dust extinction in order to derive accurate rates. For Type Ia SNe, determining the rate offers a way to estimate the delay time between the formation of the progenitor star and its explosion as an SN, which may shed light on the still mostly unknown scenarios leading up to the explosion of the degenerate white dwarfs that are assumed to be the progenitors of Type Ia SNe (e.g., Madau et al. 1998; Dahlen \& Fransson 1999; Gal-Yam \& Maoz 2004; Maoz \& Gal-Yam 2004).

Due to their potential as standard candles used to measure the expansion rate of the universe (Riess et al. 1998; Perlmutter et al. 1999; Astier et al. 2006), a large number of surveys aiming at detecting Type Ia SNe have been conducted during the last decade, resulting in a multitude of rate estimates. For a recent compilation of rates from the literature, see Graur et al. (2011).

While these results show a fair consensus that the Type Ia SNR increased by at least a factor of $\sim$3 between redshift $z=0$~and $z\sim1$, there is still an uncertainty in the rates at $z\gsim 1$, reflecting the uncertainty in such searches, the limited statistics and relatively high systematic errors concerning searches at $z>1$. Moreover, optical searches are also hampered by the redshifting of the SN spectral energy distribution (SED), making them drop out of optical filters at $z\gsim$1. Deep surveys in the near-infrared (IR), such as the {\it Hubble Space Telescope (HST)} based CANDELS (Grogin et al. 2011; Koekemoer et al. 2011) and CLASH (Postman et al. 2012) programs, should help alleviate this scarcity of high-redshift Type Ia SNe. Rodney et al. (2012) demonstrate the possibility of detecting and follow-up of Type Ia SNe at $z>1.5$~using IR searches.

While Type Ia SNe have been the focus of a large number of surveys during the last decade, there has been a lack of dedicated surveys aiming at finding CC SNe at cosmological distances. Most of the existing surveys suffer from severe selection biases since the foremost aim has been to select clean samples of Type Ia SNe and little follow-up has been devoted to non-Type Ia SN transients, including CC SNe. Furthermore, CC SNe are harder to detect and follow on the light curve because they are typically $\sim$2 mag fainter than Type Ia SNe. Therefore, there are significantly fewer CC SNRs reported in the literature compared to Type Ia SNRs.

The local CC SNR (outer distances between 11 and $\sim$200 Mpc) has been determined by Cappellaro et al. (1999), Smartt et al. (2009), Li et al. (2011), Botticella et al. (2012), and Mattila et al. (2012), while at cosmological distances rates have been determined at $z=0.21$~by Botticella et al. (2008), $z=0.26$~by Cappellaro et al. (2005; these rates are, however, superseded by the rates in Botticella et al. 2008), $z=0.3$~by Bazin et al. (2009), $z=0.66$~by Graur et al. (2011), $z=0.39$~and $z=0.73$~by Dahlen et al. (2004), and most recently at $z=0.39$~and $z=0.73$~by Melinder et al. (2012).

Most rates are consistent with each other and they increase with look-back time, reflecting the increased SFR between $z=0$~and $z=1$. However, there are some significant differences in the measured local rates. Furthermore, there are also claims by Horiuchi et al. (2011) that the SNR in general is a factor of $\sim$2~lower than predicted from the SFRs, suggesting that there are possible unaccounted systematic effects that may affect the determined rates. Most importantly, necessary corrections to account for the dimming of the SNe by dust in the host galaxies are highly uncertain. The effect of dust extinction is also expected to increase with redshift since the rest frame probed in optical searches approaches the UV part of the spectrum where the effect of extinction is more severe. There should also be a population of CC SNe almost completely hidden from optical searches in highly extinguished luminous infrared galaxies (LIRGs) and ultra-luminous infrared galaxies (ULIRGs; see, e.g.,  Mannucci et al. 2007; Mattila et al. 2004, 2012). Since the fraction of the star formation that occurs in U/LIRGs increases significantly with look-back time and actually dominates the total star formation at $z\sim 1-2$~ (e.g., Magnelli et al. 2011), we expect the number of CC SNe hidden in these environments to increase with redshift. The CC SNe hidden in such galaxies must be accounted for when the aim is to derive the complete rates of CC SNe.

Also, having a good knowledge of how the SN population is divided into different subtypes and the luminosity function of each subtype is essential for deriving accurate rates. In particular, knowing the fraction of faint CC SNe ($M\gsim$-15) is important when deriving rates since these will remain mostly undetected and must be corrected for when estimating rates.

In this paper, we use the SN sample detected in the extended GOODS/PANS SN survey conducted with $HST$/ACS 2002--2005 to derive the rate of CC SNe to redshift $z\sim$1.1. In a previous paper (Dahlen et al. 2004), we presented CC SNR to $z$=0.7 based on a subsample from the same search ($\sim$36\% of the search that was available at that time). The present paper accompanies Dahlen et al. (2008), where we presented the Type Ia SNR to $z\sim$1.6 based on the full sample. This paper is organized as follows. In Section 2, we describe the SN search and present the sample. In Section 3, we describe how SNRs are calculated. The results are presented in Section 4 together with a discussion in Section 5. We summarize our results in section 6. Throughout this paper we assume a cosmology with $\Omega_{\rm M}$=0.3, $\Omega_{\Lambda}$=0.7, and $h$=0.7. Magnitudes are given in the Vega system.     
\section{Data}
\subsection{Observations}
\begin{figure}
\epsscale{0.7}
\plotone{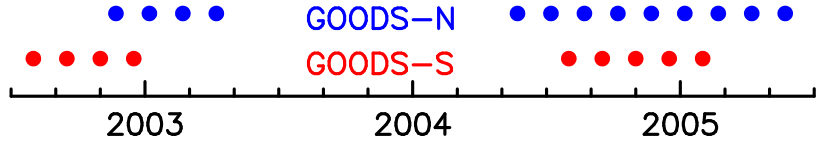}
\figcaption[f1.eps]{Cadence of the 13 search epochs in GOODS-North and 9 search epochs in GOODS-South.
\label{fig1}}
\end{figure}

The data used in this investigation were obtained during $HST$~Cycles 11--13 (2002--2005) and consist of 1 reference and 13 search images in the GOODS-North and 1 reference and 9 search images in the GOODS-South fields (Giavalisco et al. 2004; Riess et al. 2004, 2007). Both fields were observed during two separate sequences with a cadence between epochs of $\sim$45 days within each sequence. The spacing in time of the search epochs is illustrated in Figure \ref{fig1}. Each epoch consists of $\sim$15 pointings with the F850LP filter using the Advanced Camera for Surveys (ACS) on board the $HST$. The field of view of ACS is $\sim$11 arcmin$^2$, and therefore each search covers $\sim$160 arcmin$^2$. With a total of 22 epochs, we have searched an area equivalent to 1 deg$^2$ with a time separation of 45 days on average.

Besides the search filter, F850LP, we also obtained additional photometry in the F775W filter, and in Cycle 11 also in the F606W filter, which allowed us to obtain color information of the SN candidates. For a number of SN candidates we also obtained spectroscopic follow-up using both $HST$~and ground-based facilities. For more information on the observations and reduction techniques used here, see Giavalisco et al. (2004) and Strolger et al. (2004).
\subsection{The Supernova Sample}
In total, 118 SNe were detected during the full search campaign. Based on SN spectra, colors, light curves, SNe and host galaxy redshifts (spectroscopic when available, otherwise photometric redshifts based on optical+NIR bands), we classified the candidates into 56 Type Ia SNe and 62 CC SNe (see Strolger et al. 2004 for details regarding the typing of the SNe). For the CC SNe, which are the focus of this investigation, we have spectroscopic redshifts of the SN or the host galaxy for 53 SNe, while we use photometric redshifts for 6 host galaxies. The remaining three SNe are apparently hostless and for these we estimate the redshift based on the SN magnitudes, colors, and light curve. The number of CC SNe used to calculate the rates is 45, excluding from the full sample the SNe that are outside the redshift range investigated, $0.1<z<1.3$, are fainter than our adopted cut-off magnitude $m_z$=25.8 (see Section 3.2), or were detected as decliners present already in the reference epoch. 
     
\section{Calculating rates}
The SNR is calculated using Monte Carlo simulations as described in detail in Dahlen et al. (2004). In summary, the simulations assume a given input rate within a particular redshift bin and calculates the number of detectable SNe for a search with an observational setup identical to the $HST$~search, i.e., field of view, temporal spacing between epochs, number of epochs, and detection limits. A number of properties characterizing the SN population go into the simulations, such as the distribution of SN peak brightness, light curve shapes, spectral evolution, the relative fraction of different subtypes, and extinction. Rates are thereafter calculated by adjusting the input rate so that the predicted number of detectable SNe matches the number found in the actual search within each redshift bin. Below we give a brief summary of the assumptions used in this investigation when calculating the rates. 

\subsection{CC SN Subtypes} 
CC SNe are traditionally divided into different subtypes based on spectral features or light curve shapes. Since each type has different peak luminosity distributions and temporal evolution, it is important to correctly characterize each subtype as well as determine the relative frequency of each type.

\subsubsection{Peak Magnitudes}
We divide the CC SNe into four main types; Type II Plateau (IIP), Type II Linear (IIL), Type II with narrow emission lines (IIn), and Stripped Envelope SNe (SE), where the latter includes types Ib, Ic, and IIb. Furthermore, Richardson et al. (2002, 2006) note that the IIL and SE types have indications of a bi-modality in the luminosity function and present peak magnitudes for both a normal and a bright subpopulation. Here we adopt this subdivision into two different subtypes for IIL and SE types. Peak magnitudes and dispersion for the six types used here are taken from Richardson et al. (2002, 2006) and are given in Table \ref{tab1}.

\subsubsection{Subtype fractions}
To estimate the relative intrinsic fractions of the different subtypes we use the results from Smartt et al. (2009) and Li et al. (2011). Smartt et al. present the relative fractions in a volume-limited sample ($<$28 Mpc), in total consisting of 92 CC SNe, while Li et al. present a volume-limited sample ($<$60 Mpc) consisting of 133 CC SNe. Here we combine these results to improve the statistics when deriving the fractions. Note that there is an overlap of 24 SNe in the two samples, and we take this into account when combining the two. The results are presented in Table \ref{tab1}. The IIP type is most common (54.8\%) followed by the SE SNe (34.0\%). Only minor fractions belong to the IIL type (6.1\%)  and IIn type (5.1\%). However, since that latter two types are more luminous on average than the former two, particularly IIn, the observed fraction of SNe in a magnitude-limited search will likely have higher fractions of these subtypes.

To estimate the subfraction within the IIL and SE types that belong to the bright population, we use the number of bright and normal SNe in each of the samples used in Richardson et al. (2002, 2006). However, since these are basically magnitude-limited samples, we cannot use the observed fraction of bright SNe since they will be over-represented due to their brightness, which could bias the calculation of relative fractions. For the SE type, the sample in Richardson et al. (2006) consists of 4 bright and 23 normal SNe. Inspecting the distance moduli for these SNe reveals that three of the four bright SE SNe are at larger distances compared to the normal type, while one of them is at a distance where normal SNe are also detected. This suggests that 1/24, corresponding to $\sim$4\% of the SE SNe, belong to the bright group, although with high uncertainty. For the IIL subtype, 4 objects out of a total of 16 belong to the bright subsample in Richardson et al. (2002). Inspecting the distance moduli for these objects does not show any significant difference between the bright and normal subsamples. We therefore assume that 4/16, or 25\% of the IIL SNe, belong to the brighter population. In the fifth column of Table \ref{tab1} we also include this subdivision of the SE and IIL types. We again note that these fractions are uncertain. Furthermore, while these are calculated using a local sample, it is possible that the relative fraction of different types evolves with look-back time due to, e.g., a change in metallicity with time. This adds to the uncertainty. In Section 4 where we discuss systematic errors, we investigate how the uncertainty in the fractions may affect the derived rates.
\subsubsection{Fraction of faint M $>-$15 SNe}
Faint CC SNe with magnitudes $M>-$15 may be too faint to be detected even in local surveys such as the Lick Observatory Supernova Search (LOSS; Li et al. 2011), which goes out to $\sim$ 200 Mpc.  However, if the intrinsic fraction of faint SNe could be determined, then it would be possible to account for this population when deriving the rates. Richardson et al. (2002) argue that the fraction of faint $M_{\rm B}>-$15 CC SNe should be at least 20\%, perhaps much higher. This is consistent with the results of Cappellaro et al. (1997), who estimate that faint 1987A-like SNe should be 10\%--30\% of the Type II SNe. As a conservative limit on the fraction of faint SNe, we adopt a value of 20\%, consistent with the estimate of Richardson et al. (2002). The values listed in Table \ref{tab1} (Columns 2--4) already result in $\sim$4\% SNe with $M_{\rm B}>-$15 when taking into account the dispersions in peak magnitudes. To account for a larger fraction of faint CC SNe, we add a population of SNe with peak magnitude  $M_{\rm B}=-$14.4, similar to the faint SN 1987A. In order to make the total fraction of faint SNe 20\%, we assume that ~19\% of the SN population is of the faint SN 1987A type. These fractions are shown in Column 6 of Table \ref{tab1}.
\subsubsection{Light Curve and SED}
Each of the four main CC SN types is represented by a light curve and a set of SEDs that changes with time on the light curve. We base our light curves and spectral library on the templates publicly available at the Web site of P. Nugent.\footnote{http://supernova.lbl.gov/$\sim$nugent/nugent{\_}templates.html}

\subsection{Detection efficiency}
\begin{figure}
\epsscale{0.5}
\plotone{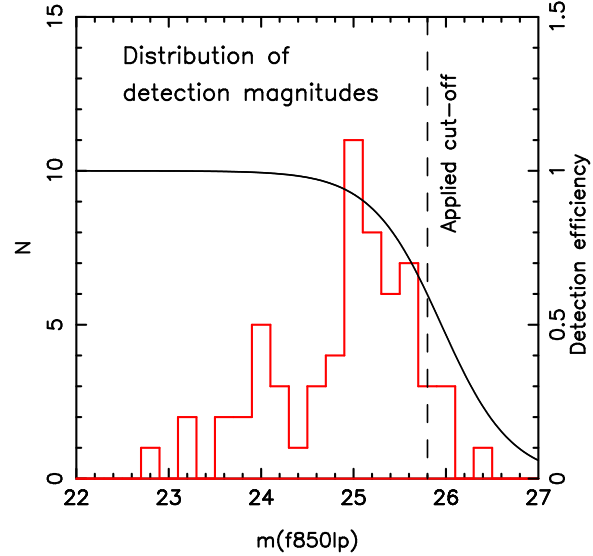}
\figcaption[f2.eps]{Distribution of detection magnitudes for the full sample of 62 CC SNe found in the $HST$~search. The solid line shows the detection efficiency derived by adding fake SNe to the search images. The vertical dashed line shows the magnitude cutoff applied when calculating rates.
\label{fig2}}
\end{figure}

The detection efficiency of the $HST$~SN search is described in Strolger et al. (2004) and Dahlen et al. (2008). In short, two methods were used to derive the detection efficiency. First, fake SNe of different magnitudes were added to the actual images during the real-time manual searches. The number of added fake SNe was purposely limited to minimize fatigue by the people doing the searches. Second, we used a Monte Carlo method to plant thousands of false point-spread functions in the data and thereafter tried to recover them using an automated detection algorithm. The Monte Carlo test can probe a significantly larger parameter space (combinations of magnitude and host environments) and is therefore complimentary to the real-time test. The Monte Carlo method makes assumptions on the limitations of the manual searches, i.e., setting a threshold for detection that matches the human searches. It is therefore not surprising that the two methods produce similar results in detection efficiency. The results from these two methods were averaged to give a final detection efficiency. The solid line in Figure \ref{fig2} shows the derived detection efficiency. The vertical dashed line shows the cutoff magnitude used when deriving rates. By applying a cutoff we decrease the need for corrections due to incompleteness at the faintest magnitudes. Even though this cutoff excludes seven SNe, the increase in error due to the lower statistics is less than the uncertainties introduced by the incompleteness corrections. Also, these are the faintest SNe in our sample and may therefore have larger uncertainties in their type determination. 

\subsection{Dust extinction}
Extinction by dust in their host galaxies makes SNe fainter and a number of SNe which without extinction would be readily detectable may fall below the detection limit of the survey. Even if the peak magnitude is above the detection limit, a fainter SN spends a shorter time on the light curve above the limit and is therefore easier to miss during a search. Correcting for the SNe missed due to these extinction effects is essential for deriving accurate rates. 
The model used in this investigation to calculate corrections for dust extinction in normal spiral galaxies is based on the models in Riello \& Patat (2005). The model assumes different exponential dust and SN distributions in galaxies. The effects of the spiral arm structure are considered to be negligible and the SNe are assumed to be found in host galaxies with random inclinations. Furthermore, the model requires the specification of a bulge-to-total (B/T) ratio. Here we use B/T = 0.
The resulting distribution of extinction values can be normalized through the face-on central optical depth through the model disk galaxy, $\tau_V(0)$. Mattila et al. (2012) show that $\tau_V(0)$=2.5 gives a good fit to observed values of host galaxy extinctions within 12 Mpc. To further check this, we add the sample in Richardson et al. (2006) to the observed sample of host galaxy extinctions in Mattila et al. (2012). We find that when using the same criteria as Mattila et al. (including the observed extinctions in galaxies with an inclination less than 60$^{\circ}$, and excluding two observed SNe with high extinction, $A_V>3$) the mean extinction is $\langle A_V\rangle=$ 0.42$\pm0.09$~for our sample of 34 SNe. This is consistent with the predicted value $\langle A_V\rangle =$ 0.44 from the simulations, after using the same inclination restriction and excluding a few objects ($\sim$1\%)  with $A_V>3$~in the high extinction tail of the simulated distribution.

At low redshift, most star formation, and therefore also CC SN production, occurs in galaxies with moderate amounts of dust extinction which we assume can be mostly corrected for by the models described above. We note, however, that while these models do predict a fairly significant tail of high extinction values in inclined host galaxies, they only result in about $\sim 1\%$ with high extinction values $A_V>3$~in galaxies with an inclination less than 60$^{\circ}$. Based on observed data, Mattila et al. (2012) estimate that a fraction of about $\sim$15\% may have such high extinction even in these low to intermediate inclination galaxies and are therefore more likely to be missed in searches. 

Furthermore, the fraction of star formation taking place in highly dust-enshrouded environments, such as LIRGs and ULIRGs, increases rapidly with redshift (e.g., P\'{e}rez-Gonz\'{a}lez et al. 2005; Le Floc'h et al. 2005; Caputi et al. 2007; Magnelli et al. 2009, 2011). For example, the results of Magnelli et al. (2011) show that at low redshift $\lsim$10\% of the star formation occurs in U/LIRGs. This increases to $\sim$50\% at $z=1$. A majority of the SNe in these environments will suffer from high extinction and should be mostly invisible in optical passbands (Mannucci et al. 2007; Mattila et al. 2012). 

To account for the SNe that are lost by optical searches due to heavy extinction in dusty starbursts, we use the results from Mattila et al. (2012), who calculate the missing fraction using an approach somewhat similar to Mannucci et al. (2007). First, the fraction of CC SNe that is hidden in highly dust-enshrouded environments within normal galaxies, LIRGs, and ULIRGs is estimated. Thereafter the fraction of the total star formation occurring in these different galaxy types and their relative contributions as a function of redshift are taken from Magnelli et al. (2011) to estimate the missing fraction of CC SNe as a function of look-back time. Using the results in Mattila et al. (2012), we find that the missing fraction increases from $\sim$19\% at low redshift to $\sim$38\% at $z\gsim$1. To take into account the redshift dependence of the missing fraction, in each of our redshift bins  we calculate the volume-averaged missing fraction using the results in Mattila et al. (2012). In order to correct the rates for the missing fraction, results should be multiplied by a de-bias factor that can be derived from the missing fraction according to the de-bias factor=1/(1-missing fraction). 

\section{Results}
\begin{figure}
\epsscale{0.5}
\plotone{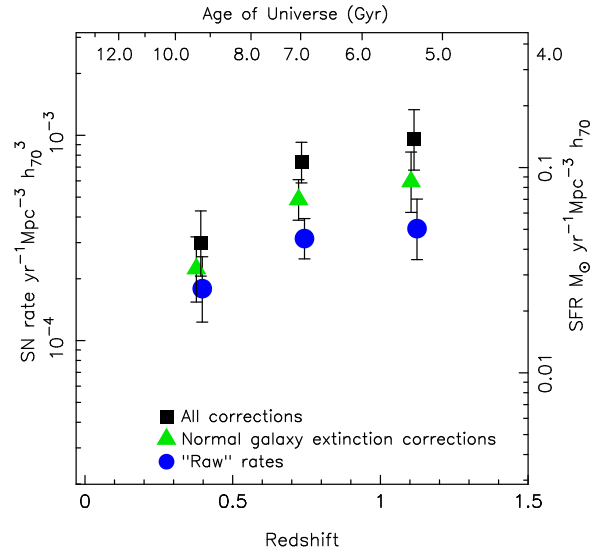}
\figcaption[f3.eps]{Core collapse SNR. The circles show ``raw'' rates without any correction for extinction or ``missing fraction''. The triangles show rates after correcting for extinction in normal galaxies. The squares show our final rates after also correcting for the fraction of CC SNe expected to be missed in dust-enshrouded environments, in particular in LIRGs and ULIRGs.
\label{fig3}}
\end{figure}

The CC SNRs are calculated in three redshift bins $0.1<z<0.5$, $0.5<z<0.9$, and $0.9<z<1.3$, centered at $z\sim$0.39, $z\sim$0.73, and $z\sim$1.11, respectively. For these bins, we find rates of 3.00$^{+1.28}_{-0.94}$, 7.39$^{+1.86}_{-1.52}$, and 9.57$^{+3.76}_{-2.80}$~in units ${\rm yr}^{-1}{\rm Mpc}^{-3}10^{-4}h^3_{70}$. The errors are statistical only, and systematic errors are discussed in the next subsection. The results are also presented in Table \ref{tab2} and are plotted in Figure \ref{fig3}. To emphasize how the dust corrections affect the results, we also give the raw rates without any corrections, the rates after applying extinction correction in normal galaxies, and finally also the rate after de-biasing for the missing fraction in high-dust environments. It is clear that the importance of the corrections due to extinction effects increases with redshift. In the lowest redshift bin, the final rates are a factor of 1.7 higher than the raw rates. This increases to factors 2.4 and 2.7 in the intermediate- and high-redshift bins, respectively.

The rates calculated for 8 of the 22 search epochs used here were reported in Dahlen et al. (2004). The main difference between the investigations is the better statistics in the current results and the implementation of the de-bias factor for missing SNe. For the $0.1<z<0.5$~bin we found in the earlier eight epoch survey a rate  2.51$^{+0.88}_{-0.75}$, which is fully consistent with the new results within statistical errors, both with and without de-biasing. For the $0.5<z<0.9$~bin, we found a rate of 3.96$^{+1.03}_{-1.06}$~${\rm yr}^{-1}{\rm Mpc}^{-3}10^{-4}h^3_{70}$~in the earlier search. These measurements are consistent within 1$\sigma$~statistical errors with the new measurements before applying the de-bias factor. After including the de-bias correction, the measurements are just outside the 1$\sigma$~statistical errors, leaving the rates in the 2004 paper which were not corrected statistically lower than the new rates. However, adding systematic errors in quadrature to the statistical errors makes the two measurement in this bin consistent within the 1$\sigma$~errors. For the highest redshift bin, $0.9<z<1.3$, there were not sufficient data on the 2004 sample to derive a rate. This comparison highlights the importance of systematic errors when deriving rates since these can give a significant contribution to the total error budget of the derived rates. In the following, we look more closely at the various systematic errors affecting the derived SNRs.

\subsection{Systematic Errors}
There are a number of possible sources that may introduce systematic errors in SNR estimates. Here we estimate how different effects may affect the derived values. The results are summarized in Table \ref{tab3}. 

\paragraph{Subtype fractions}
Since the SN subtypes have different peak magnitudes, dispersion, light curves, and SEDs, the division into subtypes will affect the ``control time'' and therefore the derived rates. In this paper, we use a division into subtypes based on a combination of the results from Smartt et al. (2009) and Li et al. (2011). To test the sensitivity to the subtype division, we recalculate the rates using the Smartt et al. and Li et al. results separately. We find that this changes rates by approximately $\sim$1\%, $\sim$3\%, and $\sim$13\% in the three redshift bins, respectively. In addition, to account for the uncertainty in the fraction of bright IIL and SE SNe, we also calculate rates after assuming that these fractions vary with $\pm$50\%. This introduces a change in the rates by $\sim$1\%, $\sim$2\%, and $\sim$7\%, respectively. In Table \ref{tab3}, we have added these two sources of uncertainty.   

\paragraph{Faint  M $>-$15 fraction}
Based on the discussion in Section 3.1.3, we assume that 20\% of the CC SNe are faint with $M>-$15. To investigate the dependence on the faint fraction, we also derive rates after assuming that 10\% and 30\% of the SNe belong to the faint population, respectively. This changes the rates by 10\%--19\%.

\paragraph{Peak magnitudes}
We take into account the dispersion on peak magnitudes as given in Table \ref{tab1} when deriving rates. Nevertheless, there are also uncertainties in the peak magnitudes which may affect the derived rates. Here we use the estimated peak magnitude uncertainties in Richardson et al. (2002, 2006) in MC simulations to quantify the systematic errors. We find that rates are affected by 4\%--10\% due to the uncertainty in peak magnitude values.

\paragraph{Redshift uncertainty}
Wrongly determined redshifts for the SN may shift objects in or out of redshift bins and therefore cause systematic errors in the derived rates. Fortunately, we have spectroscopically determined redshifts of the SN or the host for 39 of the total 45 SNe in our sample used to derive rates. We therefore do not expect that this effect should be large. Nevertheless, we use MC simulations to redistribute the redshifts for the objects without spectroscopic redshifts by using an error distribution based on the photometric redshift fitting. We find that the relative errors due to the redshift uncertainty is only a few percent, significantly less than the statistical errors. Details are given in Table \ref{tab3}.

\paragraph{Type determination}
For our full sample of 118 SNe, we have spectroscopic confirmation of the type for 2 CC SNe and 30 Type Ia SNe. For the remaining, their available photometry in the  $z$, $i$, and $V$~filters (including non-detections) was fit to model panchromatic light curves (made from low redshift analogs), redshifted and $k$-corrected to the observed passbands. The peak magnitudes and host extinction were left as free parameters in the fit, although constrained by the adopted subtype fractions (Section 3.1.2) and extinction distributions (Section 3.3). The quality of the light curve fit dictates the certainty in the SN type, which we broadly quantify as bronze, silver, or gold (in increasing certainty), following a prescription described in Strolger et al. (2004).

Melinder et al. (2011) use MC simulations to estimate the fraction of misclassified SNe for a sample with the requirement that the objects are detected in at least two epochs and that color information exists for both of these epochs. They find that 5\%--10\% of the SNe may be misclassified. In this investigation, we similarly require that the objects are detected during at least two epochs and that color information is available for both epochs. While the Melinder et al. (2012) investigation uses $R$~and $I$~as detection bands and we use $i$~and $z$~bands, we note that both investigations probe similar rest-frame bands ($B$~and $V$) at the mean redshifts for the two surveys, respectively. Based on this, it is reasonable that the misclassification rate should be similar for the two surveys. However, while the simulations assume that only photometric redshift information is available, we have spectroscopic redshifts for a majority of the objects or hosts, which should improve type determination. Furthermore, for a fraction of our sample, we have an additional color available, which also should improve typing. Therefore, our misclassification rate should be on the lower side of the 5\%--10\% range. To be conservative, we adopt a misclassification fraction that is in the middle of the estimated range, i.e.,  7.5\%. This number agrees with the fraction of CC SNe misclassified as Type Ia SNe in the full redshift $z<1.5$ sample in Melinder et al.  Assuming that 7.5\% could have a wrong type determination, we run MC simulations to investigate how this affects the derived rates. We find that rates are affected by 6\%--13\% depending on the redshift bin (see Table \ref{tab3}). This should not dominate over statistical errors.

\paragraph{Extinction corrections}
We use results based on the models of Riello \& Patat (2005) to derive the host galaxy extinction corrections. Based on results in Mattila et al. (2012), we normalize the model galaxy extinction to $\tau_V(0)=2.5$. In Section 3.3 we showed that this results in a mean absorption $A_V=0.44$~that is consistent with the mean extinction in a sample of 34 observed CC SNe, $\langle A_V\rangle$=0.42$\pm0.09$. To account for the uncertainty of the observed mean, we renormalize the extinction distributions so that the resulting extinction distribution fits the mean $\pm 1\sigma_{\langle A_V\rangle}$~and thereafter rederive the SNRs after applying the new extinctions. We find that these changes in the host galaxy extinction distributions modify the rates by 5\%--15\%, depending on the redshift bin. Details are listed in Table \ref{tab3}.

\paragraph{Extinction laws}
To calculate the wavelength dependence of the extinction, we use the prescription from Calzetti et al. (2000) with $R_V$=4.05. To investigate how sensitive results are to the specifics of the chosen extinction law, we also derive results after changing the ratio of total to selective extinction, $R_V$~by $\Delta R_V=\pm 2$. We have also tried a Cardelli et al. (1989) law with $R_V$=3.1. Using these different scenarios, we find relative changes less than 7\%, 9\%, and 12\% in the three redshift bins, respectively. 

\paragraph{De-bias correction}
In Section 3.3 we discuss the fraction of CC SNe missed in highly obscured environments, particularly in LIRGs and ULIGSs, and the de-bias factor that has to be applied to correct for the missing fraction of SNe. Using the uncertainty in the missing fraction given in  Mattila et al. (2012), we find that the resulting rates change by 10\%--42\%, depending on redshift.\\

After summing the different contributions to the systematic errors we find that statistical errors dominate in the lowest redshift bin, while the systematic errors are similar to or larger than the statistical errors in the two higher redshift bins (see Table \ref{tab3}). This emphasizes the importance of understanding the sources of possible systematic errors. In particular, corrections due to dust extinction and the fraction of faint $M>-$15 SNe need to be better understood in order to better control systematic errors.

\section{Discussion}
\subsection{Comparison to rates from the literature}
A compilation of rates from the literature is plotted in Figure \ref{fig4}, where we also show the SFR estimates from Horiuchi et al. (2011) and Magnelli et al. (2009). The latter has been put on the same scale by assuming a conversion factor between star formation and CC SNR that we calculate in Section 5.2. The figure clearly shows an increase in the CC SNR with look-back time, consistent with expectations from the evolution of the cosmic SFR. However, there is a fairly large discrepancy among the five presented local or low redshift rates. The highest local rate is presented in Botticella et al. (2012), followed by Mattila et al. (2012), Smartt et al. (2009), Li et al. (2011), and Cappellaro et al. (1999). We first note that the three first rates are based on the number of detected SNe within a specified radius during a temporal base line. These rates do not apply any corrections for SNe that could be missed due to high dust extinction, low intrinsic absolute magnitude of the SNe, or SNe missed due to insufficient coverage of the sky (e.g., SN exploding in areas that could not be observed due to solar avoidance or because SNe could also remain undetected due to the searches not covering their host galaxies such as low-luminosity dwarf galaxies). The quoted rates in these surveys are therefore lower limits. The rates of Mattila et al. (2012) and Botticella et al. (2012) are consistent within statistical error bars. This is not unexpected since the two investigations use similar selections: Botticella et al. use 14 SNe detected within 11 Mpc during 13 years (1998--2010), while Mattila et al. use 24 SNe detected in the range 6--15 Mpc during 12 years (2000--2011). The latter investigation excludes SNe within 6 Mpc so that it would not be biased by the local overdensity in star formation, and therefore also in SNR. The rate in Smartt et al. (2009) is based on 92 SNe within 28 Mpc detected during 10.5 years (1998--2008), i.e., it covers a volume of $\sim$16 and $\sim$7 times larger compared to the searches in Botticella et al. (2012) and Mattila et al. (2012). Similar to those investigations, the rate presented by Smartt et al. (2009) is not corrected for any detection biases and should therefore also represent a lower limit. When comparing to the more nearby rates, the Smartt et al. (2009) rate is about a factor of $\sim$1.7 lower. This could be partly due to the local over density affecting the most nearby rates. Also, the larger radius in the Smartt et al.  survey can contribute to this decrease due to increased distance modulus, e.g., an SN at $r=28$~Mpc  is $\sim$1.4 mag fainter than at $r=15$~Mpc and $\sim$2.0 mag fainter than at $r=12$~Mpc, this could make intrinsically faint or dust extinguished SN drop below detection limits at larger radii. The sample of Mattila et al. (2012) between 6 and 15 Mpc includes at least two SNe with high enough extinction that they would likely have remained undetected at a larger distance. Also, even though the local universe has been well monitored during the last decade, possible systematic effects due to non-coverage of different regions of the local volume should increase with radius. 

At even larger distances, Li et al. (2011) use a sample of 440 CC SNe detected in the LOSS. The SNe in this sample, which is the largest of any samples used to derive rates, are detected to distances of $\sim$200 Mpc. While the previously described rates consist of compilations of SNe detected over the full sky from various sources, the LOSS is a dedicated survey where rates are derived from imaging a sample of $\sim$10,000 galaxies during a period of slightly more than 10 years. The rates are first calculated normalized to the surveyed galaxies $K$-band luminosity in units SNuK (where 1 SNuK corresponds to one SN per 100 yr per 10$^{10}L_{\odot}$(K)). To decrease the selection effects against SNe in highly inclined hosts, galaxies with an inclination $>$75$^{\circ}$~are not included  when calculating the normalized rates. After deriving the normalized rate, the volumetric rate is derived by multiplying by the $K$-band luminosity density derived independently by Kochanek et al. (2001). When doing this, Li et al. take into account that both the normalized rates and the luminosity function differ between early-type and late-type galaxies. Furthermore, they also take into account the rate--size relation, which suggests that the normalized rates are related to the luminosity or the mass of the host galaxy. While the rate derived by Li et al. (2011) is outside the statistical errors of the rates derived at smaller distances, we do note that the risk of missing SNe due to intrinsic faintness or dust extinction, should increase with distance, possibly biasing rates to lower numbers. Also, Li et al. point out that the lack of corrections due to inclination effects (except excluding galaxies with inclination $>$75$^{\circ}$) leads to the largest uncertainty in their rates, possibly underestimating the results. There is also an additional possible systematic effect in that the rate calculation requires the use of the local $K$-band luminosity density.  

The last local rate is from Cappellaro et al. (1999) and is built on a compilation of different surveys, many of which are based on photographic plates, but also the visual search by Robert Evans (Evans 1997). The rates are calculated normalized to the $B$-band luminosity and volumetric rates are derived by multiplying by the local $B$-band luminosity density (Cappellaro et al. 2005), different from Li et al. (2011), who use the $K$-band. Considering the different search techniques, it is almost surprising that the derived rate is so close to the rates derived from CCDs. However, it is hard to evaluate different possible systematic effects which may affect the derived rates. This is outside the scope of this paper, but we do leave the rate in this paper because it is one of the first based on a compilation consisting of a significant sample of SNe.

When looking at the local rates, it is obvious that there is a trend of decreasing rates as the samples are selected at increasing radii, with deviations between some of the rates being larger than statistical errors. The trend suggests that more SNe may be missed at larger radii, possibly due to extinction in the host galaxies, which, if uncorrected for, may underestimate the rate. There could also be a relatively high fraction of intrinsically faint SNe that would be more easily missed at large radii, which, if not accounted for, would underestimate the derived rates. 

One should also note that there is a local overdensity in star formation in the very nearby universe, which could drive the most local rates to high values. Assuming that the Botticella et al. (2012) rate, measured within 11 Mpc, is affected by the local over density with approximately a factor of 1.5, which is the ratio of the nearby SFR within 11 Mpc from 11HUGS (Kennicutt et al. 2008; Bothwell et al. 2011) and a mean of a sample local measurement at larger radii going out to $\gsim$100 Mpc reported in Horiuchi et al. (2011), we find that after correction the Botticella et al. rate is still larger than the rate from Smartt et al. (2009) et al. and Li et al. (2011), even though it still represents a lower limit. On the other hand, Botticella et al. (2012) note that when normalizing their rate to the local $B$-band luminosity density, they find a value consistent with the normalized rate at $z\sim$0.01 in Cappellaro et al. (1999), suggesting that local overdensity can fully explain their elevated local rate. Moreover, Mattila et al. (2012) noted that at very small radii, $r\lsim$6 Mpc, there is an elevated SNR, reflecting the over abundance in the SFR in the nearby universe. They show that the rate calculated within $r<$6 Mpc is about three times the rate at $r<$10 Mpc. To avoid being affected by the high number of SNe found at small radii, the rate in Mattila et al. is based on SNe found in the range 6 Mpc$<$~radius $<$15 Mpc, which should limit the influence of a local overdensity on their rates. Still, it cannot be ruled out that the rate in Mattila et al. is elevated due to the effects of the local overdensity.

A non-local rate at $z\sim0.21$~based on $\sim$45 CC SNe is presented in Botticella et al. (2008). The SNe are from the Southern inTermediate Redshift ESO Supernova Search (STRESS) and the rates replace the preliminary results from a subsample of the survey given in Cappellaro et al. (2005). While this is consistent with the local rates in Botticella et al. (2012) and Mattila et al. (2012), there is an indication that the rate is on the low side to be consistent with the expected increase in the SFR, and therefore also SNR with look-back time. We note that Botticella et al. (2008) correct for extinction in normal galaxies, but do not make any correction for U/LIRG like environments, which could have some effect on the rates. There could also be a number of systematic effects affecting the rate. This includes the fact that only about one-third of the transients in the survey has an SN-type determination and that for another third there is an ambiguity whether the transient is an SN or an active galactic nucleus. Furthermore, this was a pointed survey, observing a sample of galaxies, with the aim at deriving the $B$-band luminosity-normalized SNR. Deriving the volumetric rate based on this includes the extra systematic uncertainty when multiplying the luminosity-normalized rate with the global luminosity density at the redshift range of interest.

At somewhat higher redshift $z\sim$0.3, Bazin et al. (2009) present the CC SNR, based on a sample of 117 SNe from the Supernova Legacy Survey, which is the largest non-local sample of SNe used to derive rates. Similar to the Botticella et al. (2008) rate, this rate is consistent with lower redshift rates; however, it does not show an increase compared to the most recently published local rates, which would be expected if the rate increases with look-back time. Again, we note that while the Bazin et al. (2009) rate includes extinction corrections in normal host galaxies, it does not correct for highly dust extinguished regions, which could somewhat bias the rates to lower values. Furthermore, there are systematic effects that are difficult to quantify in the derived CC SNR. In particular, Bazin et al. (2009) do not directly derive the CC SNR; instead they estimate the relative rate between CC SNe and Type Ia SNe at $z\sim$0.3 and then use the Type Ia SNR derived at $z\sim$0.5 by Neill et al. (2006), which is extrapolated to $z\sim$0.3, and thereafter multiplied by the assumed ratio to derive the CC SNR. 

At higher redshifts, overlapping the range presented in this paper, Graur et al. (2011) derive the rate at $z\sim$0.66 based on three SNe classified as CC SNe, while Melinder et al. (2012)  present rates at $z\sim$0.39 and $z\sim$0.73 using three and five CC SNe in the two bins, respectively. Both of these results apply corrections for host galaxy dust extinction as well as galaxies missing in U/LIRGs, using methods similar to those used here. As can be seen in Figure \ref{fig4}, these rates are consistent with the rates derived here in the overlapping redshift range. At the same time, the error bars on the Graur et al. and Melinder et al. rates are large, covering a factor 2--10 in the derived rates, mainly due to the low SN statistics. Additional systematic errors could also be large in those results due to, e.g., the lack of spectroscopic confirmation of the redshifts of both the SNe and the host galaxies. With our new rates, based on the largest high-redshift statistical sample for which we have spectroscopic confirmation of the redshift of the SN or host for a majority of the sample, we are able to significantly decrease the size of the error bars, making it possible to set firmer constraints on the evolution of the cosmic SNR.

\subsection{The star formation rate connection}
One of the most important diagnostics that the CC SNe can be used for is the measurement of the cosmic SFR. Note that the CC SNR provides a completely independent method for measuring the evolution of the SFR, compared to other widely used methods based on, e.g., galaxy luminosity densities (UV, IR, H$\alpha$, etc.). For a direct comparison between the local CC SNR and different SFR indicators, see Botticella et al. (2012). Previous results have shown that the increase in CC SNR with look-back time is consistent with an increase in the rate as expected from the SFR (Dahlen et el. 2004). However, the details in the evolution of the CC SNR and their normalization to the SFR still have large uncertainties, which has prevented a more detailed comparison. 

\begin{figure}
\epsscale{0.5}
\plotone{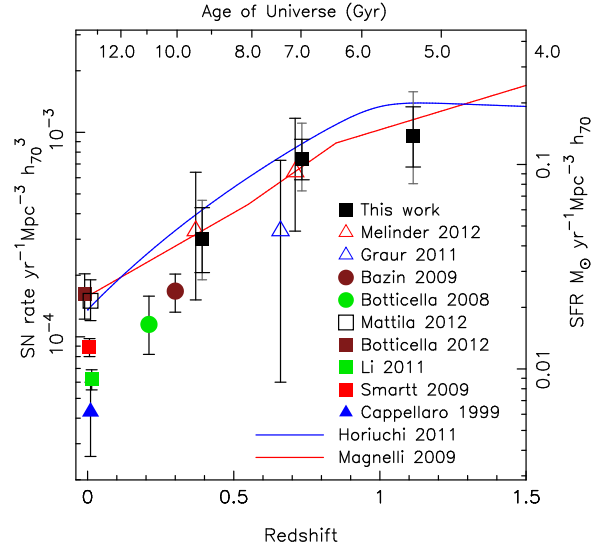}
\figcaption[f4.eps]{Core collapse SNRs from this work together with rates from the literature. Shown also are the star formation rates from Magnelli et al. (2009) and Horiuchi et al. (2011) after assuming a relation between star formation and CC SNR given in Section 5.2. Rates from the literature are shown with statistical errors. For the rates in this work, we show statistical errors with black error bars, while gray (larger) error bars show the added statistical and systematic errors (in quadrature). Note that the local rates within 11--15 Mpc by Botticella et al. (2012) and Mattila et al. (2012) may be elevated due to the local overdensity in star formation. See Section 5.1 for details.
\label{fig4}}
\end{figure}

In Figure \ref{fig4}, we plot the CC SNR with the SFR from Magnelli et al. (2009) and Horiuchi et al. (2011). To relate the CC SNR and the SFR, we assume a Salpeter initial mass function (IMF; Salpeter 1955), $\psi(M)$, in the range $0.1<M/M_{\odot}<125$~together with the assumption that all stars with masses in the range  $8<M/M_{\odot}<50$~explode as CC SNe. We can thereafter relate the SFR (SFR in units of $M_{\odot}$yr$^{-1}$Mpc$^{-3}$) and the CC SNR (SNR in units of yr$^{-1}$~Mpc$^{-3}$) by
\begin{equation}
{\rm SNR}(z)=k\times h^2\times {\rm SFR}(z), 
\end{equation}
where 
\begin{equation}
k = \frac{\int^{50\mathcal{M}_{\sun}}_{8\mathcal{M}_{\sun}} \psi(M) dM}{\int^{125\mathcal{M}_{\sun}}_{0.1\mathcal{M}_{\sun}} M \psi(M) dM}.
\end{equation}
Numerically, this results in $k$=0.0070 $M_{\odot}^{-1}$. There are two main uncertainties in deriving this value. First, the choice of IMF affects the result. However, the dependency of the IMF in Equation (2) is mostly cancelled out since the derived SFR in Equation (1) also depends on the IMF. E.g., Melinder et al. (2012) show that even if making a significant change in the IMF, such as choosing a modified Salpeter IMF from Baldry \& Glazebrook (2003), which will  increase $k$~by a factor of $\sim$2, the resulting change in the relation between SNR and SFR is only $\sim$2\% due to the change of the SFR normalization. The second uncertainty comes from the selected range of CC SN progenitor masses. Here we use a lower limit of 8$M_{\odot}$~consistent with the results of Smartt et al. (2009), who estimate that the minimum stellar mass for type IIP progenitors lies in the range $7<M/M_{\odot}<9.5$. This range corresponds to an uncertainty in $k$~by +22\%/--23\%. The upper limit (50$M_{\odot}$) is taken from Tsujimoto et al. (1997), who suggest that progenitors more massive than this become black holes directly instead of exploding as SNe. Due to the steep slope of the IMF, the actual choice of upper limit has limited impact on the derived $k$~value. Here we assume that the upper limit is restricted to the range $30<M/M_{\odot}<100$, resulting in an uncertainty of +6\%/--9\%. Taking these ranges in upper and lower limits into account, we find that $k$~should lie in the range 0.0048$<k<$0.0089, indicating a maximum uncertainly in the normalization by less than $\pm$25\%. We also note that there is an uncertainty in the normalization of the SFR  by approximately $\sim$20\% (e.g., Figure 5 in Horiuchi et al. 2011). An uncertainly of this order is also illustrated on Figure \ref{fig3} as the difference in SFR between the estimates in Magnelli et al. (2009) and Horiuchi et al. (2011). Therefore, there should still be a total systematic uncertainty of $\sim$32\% when comparing the SFR derived from galaxy luminosities and the SFR from CC SNe. 

Using the relation in Equation (1), we see in Figure \ref{fig4} that our newly derived rates well match the predictions from the SFR over the range $0.4\lsim z \lsim 1.1$. Indicating that the assumptions made when deriving the rates are not severely affected by systematic errors, and the assumed relation between SFR and SNR seems reasonable. We note that the local rates by Botticella et al. (2012) and Mattila et al. (2012) are also consistent with the expectations from the SFR, even though these rates represent lower limits. 

Based on the CC SNRs published prior to 2011, Horiuchi et al. (2011) noted that the observed CC SNRs and the rates expected from the SFR are systematically offset by a factor of $\sim$2, particularly at lower redshifts. The main explanations for this discrepancy suggested that a high fraction of CC SNe could be missing from the searches due to a significant population of intrinsically faint SNe or that a large fraction is invisible to searches since they are hidden behind significant amounts of dust. In this investigation, we take both of these sources into account when deriving rates. Since our results agree well with what is expected from the SFR, it seems likely that the discrepancy in Horiuchi et al. (2011) is indeed due to the factors suggested by the authors. For the other high-redshift rates, the inclusion of corrections for SNe missed in U/LIRGs in the rates of Graur et al. (2011) and Melinder et al. (2012) should contribute to increase these rates to make them consistent with the SFR. For local rates, Mattila et al. (2012) report a value that is also consistent with the SFR. They argue that within 15 Mpc, a large fraction of either faint or dust-enshrouded SNe should not be missing, which is the reason that their rate is consistent with the SFR. Similar arguments should hold for the rate in Botticella et al. (2012), even though this rate could be elevated more due to the local overdensity in star formation. For the remaining rates in Figure \ref{fig4} that fall below the SFR lines, it is likely that these two effects both contribute to the rates, in combination with unaccounted systematic errors that could affect each different survey.

We finally note that the good agreement between the SFR and the CC SNR derived here over the whole redshift range to $z\sim$1.1 indicates that we are not significantly affected by incompleteness at higher redshifts where the SNe become fainter. Specifically, this demonstrates that we are quite complete to $m$(f850lp)=25.7, typical of a CC SN at $z\sim$1.1 (with $M_B$=-17.4). Similarly, for the Type Ia SNR published in Dahlen et al. (2008), based on the same SN search, we should therefore also be complete to $m$(f850lp)=25.7, corresponding to faint SNe Ia (with $M_B$=-18.3) at redshifts $z\sim$1.6, the highest bin where the Type Ia rate was derived. This suggests that the drop in the Type Ia rate at $z>1.4$~noted in Dahlen et al. (2008) is not due to incompleteness.

\section{Conclusions and summary}
We have used a sample of 45 CC SNe, out of a total of 62 detected with ACS during the $HST$-extended SN search in Cycles 11--13, to derive the CC SNR to $z\sim$1.1. Our main conclusions are as follows. 
\begin{itemize}   
\item{After correcting for host galaxy extinction and the missing fraction in highly obscured environments, we find rates 3.00$^{+1.28}_{-0.94}$$^{+1.04}_{-0.57}$~at $\langle z\rangle$=0.39, 7.39$^{+1.86}_{-1.52}$$^{+3.20}_{-1.60}$~at $\langle z\rangle$=0.73, and 9.57$^{+3.76}_{-2.80}$$^{+4.96}_{-2.80}$~at $\langle z\rangle$=1.11. The rates are given in units of yr$^{-1}$~Mpc$^{-3}~10^{-4}~h_{70}^3$. The first errors represent statistical while the second ones are the estimated systematic errors.} 
\item{Our rates at $z\gsim$0.4 are consistent with those expected from the cosmic SFR.}
\item{Statistical uncertainties dominate our rate estimates in the low-redshift bin, however; systematic errors start to dominate at $z>0.5$.}
\item{While the most recent local CC SNRs are consistent with those expected from the SFR, there is a discrepancy between some earlier derived $z\lsim$0.3 rates and the SFR.  As noted in Horiuchi et al. (2011) and Mattila et al. (2012), this is likely due to a combination of SN missed due to dust extinction and/or faint intrinsic luminosity and unaccounted systematic effects.}
\end{itemize}
\acknowledgments{
We thank the anonymous referee for useful comments and suggestions which imroved this paper. This work is based on observations made with the NASA/ESA {\it Hubble Space Telescope}, obtained at the Space Telescope Science Institute, which is operated by the Association of Universities for Research in Astronomy, Inc., under NASA contract NAS 5-26555. These observations are associated with programs GO-9352, GO-9425, GO-9583, GO-9728, GO-10189, GO-10339, and GO-10340. S.M. and E.K. acknowledge financial support from the Academy of Finland (project: 8120503).
}

\clearpage

\clearpage
\begin{table}
\caption{Subtypes of Core Collapse Supernovae}
\centering
\begin{tabular}{llllll}
\hline
Type & Peak $M_B$ & Dispersion & Fraction$^a$ & Fraction$^b$ & Fraction$^c$\\
\hline
IIP               & --16.67 & 1.12 & 54.75\% & 54.75\% & 44.31\% \\
\hline
IIL normal        & --17.23 & 0.38 & 6.10\%  & 4.57\%  &  3.70\% \\
IIL bright        & --18.94 & 0.51 & ~       & 1.52\%  &  1.23\% \\
\hline
IIn               & --18.82 & 0.92 & 5.12\%  & 5.12\%  &  4.14\% \\
\hline
SE normal         & --17.07 & 0.49 & 34.03\% & 32.61\% & 26.24\% \\
SE bright         & --19.38 & 0.46 & ~       &  1.42\% &  1.31\% \\
\hline
Faint ($M_B{>-15}$) & --14.4  & 0.5  & ~       &        ~& 19.06\%\\
\hline
\end{tabular}
\tablecomments{ $^a$Fractions of main types derived from data in Smartt et al. (2009) and Li et al. (2011). $^b$Fractions after dividing the SE and IIL subtypes into a normal and a bright population, based on the data in Richardson et al. (2002; 2006). $^c$Fractions after assuming that in total 20\% of the SN population is fainter than $M_{\rm B}>$-15.  
}
\label{tab1}
\end{table}

\begin{table}
\caption{Type Core Collapse Supernova Rates}
\centering
\begin{tabular}{lccccccccc}
\hline
Redshift Bin & Redshift$^a$ & SNR$^b$ & Error(stat+sys) & SNR$^c$ & Error(stat+sys)& SNR$^d$ & Error(stat+sys)&$N$ & $N_{\rm dist}$\\
\hline
$0.1< z\le0.5$ & 0.39 & 1.79 & $^{+0.77}_{-0.56}$ $^{+0.30}_{-0.26}$ & 2.24 &$^{+0.96}_{-0.70}$ $^{+0.46}_{-0.37}$& {\bf 3.00} &$^{\bf +1.28}_{\bf-0.94}$ $^{\bf +1.04}_{\bf -0.57}$&9 & 9.99 \\
$0.5< z\le0.9$ & 0.73 & 3.14 & $^{+0.79}_{-0.64}$ $^{+0.70}_{-0.52}$ & 4.86 &$^{+1.22}_{-1.00}$ $^{+1.36}_{-0.92}$&{\bf 7.39} &$^{\bf +1.86}_{\bf-1.52}$ $^{\bf+3.20}_{\bf-1.60}$& 25 & 23.56\\
$0.9< z\le1.3$ & 1.11 & 3.51 & $^{+1.38}_{-1.03}$ $^{+0.82}_{-0.91}$ & 5.95 &$^{+2.34}_{-1.74}$ $^{+1.81}_{-1.60}$&{\bf 9.57} &$^{\bf+3.76}_{\bf-2.80}$ $^{\bf+4.96}_{\bf-2.80}$& 11 & 11.44\\
\hline
\end{tabular}
\tablecomments{$^a$The effective redshift is defined as the redshift that divides the volume in the redshift bin into equal halves. $^b$Rates without any corrections. $^c$Rates corrected for extinction in normal galaxies. $^d$Final rates corrected for SNe hidden in high extinction regions, particularly in U/LIRGs. Rates are given in units yr$^{-1}$Mpc$^{-3}~10^{-4}~h_{70}^3$, assuming a cosmology with $\Omega_M=0.3$ and $\Omega_{\Lambda}=0.7$. Statistical and systematic errors are shown separately. The last two columns give the number of SNe in each bin, where $N$~is the ``raw'' counts and $N_{\rm dist}$~gives the number after taking into account the redshift probability distribution of the SNe.
}
\label{tab2}
\end{table}

\begin{table}
\caption{Error sources}
\centering
\begin{tabular}{lccc}
\hline
 ~ & Redshift & Redshift & Redshift\\
\hline
~                          & 0.1$<z<$0.5 & 0.5$<z<$0.9 & 0.9$<z<$1.3 \\
\hline
Subtype fraction           & $^{+0.9\%}_{-1.4\%}$   &  $^{+3.7\%}_{-4.2\%}$ & $^{+13.9\%}_{-15.0\%}$  \\
Faint ($M>-15$) fraction &   $^{+13.5\%}_{-10.0\%}$   &  $^{+18.6\%}_{-13.2\%}$ & $^{+14.0\%}_{-13.5\%}$  \\
Peak magnitudes  & $^{+4.7\%}_{-4.0\%}$   &  $^{+8.7\%}_{-7.7\%}$ & $^{+10.5\%}_{-9.7\%}$  \\
Redshift uncertainty  & $^{+4.1\%}_{-2.7\%}$   &  $^{+1.1\%}_{-1.8\%}$ & $^{+2.5\%}_{-1.7\%}$  \\
Type determination  & $^{+7.5\%}_{-9.2\%}$   &  $^{+7.8\%}_{-6.4\%}$ & $^{+6.1\%}_{-13.2\%}$  \\
Extinction correction& $^{+9.7\%}_{-4.6\%}$   &  $^{+14.5\%}_{-6.9\%}$ & $^{+15.5\%}_{-6.8\%}$  \\
Extinction laws & $^{+6.5\%}_{-6.3\%}$   &  $^{+8.8\%}_{-4.0\%}$ & $^{+11.8\%}_{-0.2\%}$  \\
Missing fraction & $^{+28.0\%}_{-9.6\%}$   &  $^{+33.0\%}_{-10.3\%}$ & $^{+42.0\%}_{-11.6\%}$  \\
\hline
Systematic summed & $^{+34.6\%}_{-19.0\%}$   &  $^{+43.3\%}_{-21.6\%}$ & $^{+51.8\%}_{-29.3\%}$  \\
\hline
Statistical errors  & $^{+42.8\%}_{-31.2\%}$   &  $^{+25.2\%}_{-20.5\%}$ & $^{+39.3\%}_{-29.2\%}$\\
\hline
\end{tabular}
\tablecomments{ Different sources contributing to systematic uncertainties. For the summed errors, the difference sources are added in quadrature. See the text for details.
}
\label{tab3}
\end{table}

\end{document}